\documentclass[apj]{emulateapj}

\usepackage{graphicx}
\usepackage{newtxtext,newtxmath}
\usepackage{psfig}
\usepackage{natbib}
\usepackage{color}
\usepackage{soul}

\def\sn{\mbox{SN\,2014J}}

\def\asec{\ifmmode ^{\prime\prime}\else$^{\prime\prime}$\fi}

\def\msun{\hbox{M$_{\odot}$}}
\def\rsun{\hbox{R$_{\odot}$}}

\def\msunyr{\mbox{\,${\rm M_{\odot}\, yr^{-1}}$}}
\def\mdot{\dot M}

\def\degs{\ifmmode ^{\circ}\else$^{\circ}$\fi}
\def\amin{\ifmmode ^{\prime}\else$^{\prime}$\fi}
\def\asec{\ifmmode ^{\prime\prime}\else$^{\prime\prime}$\fi}

\def\degs{\ifmmode ^{\circ}\else$^{\circ}$\fi}
\def\amin{\ifmmode ^{\prime}\else$^{\prime}$\fi}

\def\EE#1{\times 10^{#1}}

\def\cm{\mbox{\,cm}}

\def\cm3{\mbox{\,cm$^{-3}$}}
\def\kms{\mbox{\,km~s$^{-1}$}}

\def\kms{\mbox{\,km s$^{-1}$}}

\def\lsim{\!\!\!\phantom{\le}\smash{\buildrel{}\over
 {\lower2.5dd\hbox{$\buildrel{\lower2dd\hbox{$\displaystyle<$}}\over
                                 \sim$}}}\,\,}
\def\gsim{\!\!\!\phantom{\ge}\smash{\buildrel{}\over
{\lower2.5dd\hbox{$\buildrel{\lower2dd\hbox{$\displaystyle>$}}\over
                               \sim$}}}\,\,}
\def\albanova{1}
\def\iaa{2}
\def\unizar{3}
\def\eso{4}

\shorttitle{Constraining magnetic field amplification in SN shocks using  radio observations of SNe 2011fe and 2014J}
\shortauthors{Kundu et al.}

\begin{document}

\title{Constraining magnetic field amplification in SN shocks using radio observations of SNe 2011fe and 2014J}

\author{E. Kundu \altaffilmark{\albanova},
P. Lundqvist \altaffilmark{\albanova}, 
M.~A. P\'erez-Torres  \altaffilmark{\iaa,\unizar},  
R. Herrero-Illana \altaffilmark{\eso}, 
A. Alberdi \altaffilmark{\iaa}
}

\altaffiltext{\albanova}{Department of Astronomy and The Oskar Klein Centre, AlbaNova, Stockholm University, SE-10691 Stockholm, Sweden. esha.kundu@astro.su.se}
\altaffiltext{\iaa}{Instituto de Astrof\'isica de Andaluc\'ia, Glorieta de las Astronom\'ia, s/n, E-18008 Granada, Spain.}
\altaffiltext{\unizar}{Visiting Scientist: Departamento de F\'isica Teorica, Facultad de Ciencias, Universidad de Zaragoza, Spain.}
\altaffiltext{\eso}{European Southern Observatory (ESO), Alonso de C\'ordova 3107, Vitacura, Casilla 19001, Santiago de Chile, Chile.}

\begin{abstract}
We modeled the radio non-detection of two Type Ia supernovae (SNe) 2011fe and 2014J considering synchrotron
emission from the interaction between SN ejecta and the circumstellar medium. For an ejecta with the outer part having a power law density structure we compare synchrotron emission with radio observations. Assuming that 20$\%$ of the bulk shock energy is being shared equally between electrons and magnetic fields we found a very low density medium around both the SNe. A less tenuous medium with particle density $\sim$ 1 $\rm cm^{-3}$, which could be expected around both SNe, can be estimated when the magnetic field amplification is less than that presumed for energy equipartition. This conclusion also holds if the progenitor of SN~2014J was a rigidly rotating white dwarf (WD) with a main sequence (MS) or red giant companion. For a He star companion, or a MS for SN~2014J, with 10$\%$ and 1$\%$ of bulk kinetic energy in magnetic fields, we obtain a mass loss rate $< 10^{-9}$ and $< \sim 4\times 10^{-9}$ \msunyr for a wind velocity of 100 \kms. The former requires a mass accretion efficiency $>$ 99$\%$ onto the WD, but is less restricted for the latter case. However, if the tenuous medium is due to a recurrent nova it is difficult from our model to predict synchrotron luminosities. Although the formation channels of SNe~2011fe and 2014J are not clear, the null detection in radio wavelengths could point toward a low amplification efficiency for magnetic fields in SN shocks.
\end{abstract}


\section{Introduction}
\label{sec:intro}
Astrophysical shocks 
are suitable places for high energy particle acceleration and magnetic field amplification.
The accelerated ions, via plasma instabilities, could amplify magnetic fields \citep{bykov13}. As particle acceleration is not very clear until now, how much magnetic field is generated in these shocks still remains an open question. Recently, \citet {cap14a,cap14b} tried to investigate these mechanisms through hybrid simulations. However, their effort is restricted by computational limitations, which do not allow to simulate a shock having a very high Mach number like supernova (SN) shocks. Another way to understand these complex phenomena is through observing radiations emitted by the shock accelerated particles. A highly relativistic particle loses a fraction of its energy in the form of synchrotron radiation. The power of this radiation is proportional to the energy density of the ambient magnetic field. Therefore, a proper modeling of synchrotron emission from these shocks along with observations could give an estimate of the ambient magnetic field, and this can then shed light on the complex mechanism of particle acceleration.  

\par
The interaction of high velocity SN ejecta with the ambient medium launches a strong shock into the circumstellar medium (CSM). Particles accelerated in these shocks can emit synchrotron radiation in radio wavelengths. When these SNe are nearby the probability of detection increases. In this regard, the two nearest young SNe~Ia, SN~2011fe, in M101, and SN~2014J, in M82, at distances of 6.4 Mpc \citep{sha11} and 3.4 Mpc \citep{mar15}, respectively, provide a unique opportunity to detect the radiation and understand the circumstances which lead to that emission. Several observational attempts were made to detect radio emission from these two SNe, but nothing was detected within the first 150 days of explosion \citep{horesh12,cho12,per14,cho16}. As reported here, observations show null detection even around 1.5 and 4 years after the explosion of SNe~2014J and 2011fe, respectively.

 \par
 The CSM plays an important role in shock dynamics. This medium is shaped by the pre-SN mass lost history of the progenitor system. Both the aforementioned SNe are of Type Ia which  are widely accepted as the thermonuclear explosion of a CO white dwarf (WD) \citep{hoyle60}. Before the explosion, this WD accretes matter from a companion star, which could be a main sequence (MS) or an asymptotic giant branch star, and explodes when it reaches the Chandrasekhar (CH) mass limit \citep{whe73}. This is called single degenerate (SD) channel. A second possibility is to have another WD as a companion with a total mass of the system being more than the CH limit. In this case, known as the double degenerate (DD) scenario, the two spiralling WDs merge, due to the emission of gravitational waves which reduces the orbital separation, and under right physical conditions give rise to a successful thermonuclear run away \citep{iben84, web84}. The companion star, in the SD scenario, loses a fraction of its initial mass either due to a wind or Roche-lobe overflow, which results in a high density circumbinary medium. On the contrary, the merger model predicts a very clean environment around the progenitor system.
 
 Previous modelings of null detection of radio emission from SNe~2011fe and 2014J \citep{cho12,per14,cho16} prefer to put stringent upper limits on the ambient medium density, $n_{\rm ISM}$, and the mass loss rate, $\mdot$, of the progenitor system assuming equipartition of energy, in the post shock region, between electric and magnetic fields. This assumption suggests a very tenuous medium around both the SNe. 
  The radio silent feature of these SNe could also be due to a lower strength of magnetic fields in these shocks. In the present work we examine both equipartition and a lower energy density in magnetic fields compared to in electric fields, and try to constrain the magnetic field amplification in SN shocks. 
  
  The paper is organised as follows. In the next section we discuss the explosion models of WDs used and the SN ejecta profile. Along with this we discuss the synchrotron radio emission model. Radio observations are given in $\S$\ref{sec:data}. We present our  results in $\S$ \ref{sec:results} and discuss them in $\S$\ref{sec:discussion}. Conclusions are drawn in $\S$\ref{sec:con}. 
  
\section{Explosion Models and Radio Emission}
\label{sec:explsnModel}
Both SNe~2011fe and 2014J are normal SNe~Ia with a $^{56}$Ni mass of $\sim 0.6$ \msun \citep{rop12,chura14}. Within the framework of the SD scenario, the N100 model \citep{sei13} is capable of reproducing the observables of normal SNe~Ia with reasonable accuracy. This is a delayed detonation model where the central region is ignited by 100 sparks. Another channel which could account for normal SNe~Ia is the merger of two sub CH WDs \citep{pak12}. Here two C/O degenerate stars of masses 1.1 \msun\  and 0.9 \msun\  merge and lead to a successful SN explosion. This
model probes the DD formation channel. The total mass and asymptotic kinetic energy of ejecta for N100 and the violent merger model are 1.4 \msun\ , 1.95 \msun\ and 1.45 $\EE{51}$ erg, 1.7 $\EE{51}$ erg, respectively.   

For both models, numerical simulations provide density of the ejecta as a function of velocity, in the homologous phase, up to $\sim 2.5\EE4 \kms$. The ejecta structure of the extreme outer part of the exploding WD is not available from these simulations. This part of the ejecta interacts with the CSM, and hence plays an important role in radio emission. Therefore, beyond the above velocity range
we have extrapolated the density ($\rho$) profile considering a power-law $\rho \propto r^{-n}$, where $r$ represents the radius of the SN and $n$ is the power-law index. The initial structure of outer part of WD is uncertain. Accretion rate and subsequent burning phases play a vital role in determining the properties of the outer edge. We therefore allow $n$ to vary in the range $12-14$\footnote{A power-law fit to the outer part of the N100 model, for velocities $\sim 2\EE4 \kms$, gives an index of around 12.}. For convenience we call the N100 + power law structure as N100 model and similarly the violent merger + power law profile as merger model.

In the SD scenario the SN ejecta most likely interact with a wind medium characterised by a wind velocity ($v_w$) and mass loss rate $\mdot$  of the progenitor system. When $\mdot$ is constant the density of this surrounding medium is expressed as $\rho_{\rm CSM} = \dot M/(4\pi r^2 v_w)$. On the other hand, in the DD formation channel, the ejecta launches a strong shock into a presumably constant density medium with $\rho_{\rm CSM} = n_{\rm ISM}\mu$ where $\mu$ represents the mean atomic wight of the surrounding medium. Therefore the density of the CSM can be written as $\rho_{\rm CSM} = A r^{-s}$ with $s = 2$ or $0$ for a wind or a constant density medium, respectively. 
The interaction of the ejecta, with power law structure, and this surrounding medium gives rise a self-similar structure \citep{che82a}. In this case, the forward shock radius, $r_s$, varies with $A$ and time, $t$, as $r_s \propto A ^{1/(s-n)} ~ t^{(n-3)/(n-s)}$ which implies a shock speed of $v_s = \frac{dr_s}{dt} = \frac{(n-3)}{(n-s)} r_s/t$. 

The synchrotron power is inversely proportional to the square of the mass of the emitting particle. Therefore, in a medium with low magnetic field strength, electrons rather than ions are more likely to be responsible for this radiation. It is expected that along with ions electrons are also accelerated in shocks. For a test particle, in a non-relativistic shock having a compression factor of 4, diffusive shock acceleration predicts a power law energy spectrum with an index $p = 2$. When synchrotron or inverse Compton losses are important the index of the integrated particle spectrum becomes $(p+1)$. 
The presence of a sub-shock along with the main shock also predicts a steep spectrum for low energy particles \citep{ellison91}. In case of SNe type Ib/c, which originate from compact progenitors like SNe type Ia, it is observed that the spectrum of relativistic particles is steeper, with an index $p = 3$ with the loss mechanisms having no role to play \citep{che06}. These loss mechanisms are found to be irrelevant for both the SNe we are concerned with here. The insufficient knowledge of particle acceleration, at shocks, does not allow a unique choice of $p$. 
We, therefore, assume that the energy spectrum of the electrons obeys $dN/dE = N_0E^{-p}$ with $p = 3$, considering the above similarities with SNe type Ib/c. Here $N_0$ and $E$ represent the normalisation constant of distribution and electron energy, respectively.
We assume that all electrons in the post shock region are accelerated. For a power-law energy distribution the minimum energy of this population varies as $E_{\rm min} \propto r_s^2 t^{-2} \epsilon_{\rm e}$ for a constant value of $\epsilon_{\rm e}$, where  $\epsilon_{\rm e} = \epsilon_{\rm nrel} + \epsilon_{\rm rel}$ = $u_{\rm e} /u_{\rm th}$ represents the fraction of post shock energy density,
$u_{\rm th} = \frac{9}{8}\rho(r) v_s^2$, that goes into electrons. $u_{\rm e}$ is the energy density in electric fields. This energy is shared by relativistic electrons with $E > m_e c^2 (\equiv E_{\rm rest})$, where $m_e$ and $c$ are the electron mass and velocity of light in vacuum, and non-relativistic electrons with a kinetic energy of the electron $E_{k} < E_{\rm rest}$. The two quantities $\epsilon_{\rm rel}$ and $\epsilon_{\rm nrel}$ give an account of the shared energy between the two components. Initially, when the shock velocity is very high, the entire electron population is relativistic. This implies that $\epsilon_{\rm e} = \epsilon_{\rm rel}$ and $\epsilon_{\rm rel} > 0.16 {\bigg(\frac{v_s}{5\EE4 ~\rm km~s^{-1}}\bigg)}^{-2}$ for $p = 3$ \citep{che06}.
 When the above criterion is not fulfilled there exists a non-relativistic electron population.
 The presence of this component implies that $\epsilon_{\rm rel} = \epsilon_{\rm e} \bigg(\frac{E_{min}}{E_{\rm rest}}\bigg)^{(p-2)}$. We consider electrons with $E > E_{\rm rest}$ to contribute to the synchrotron emission as only relativistic particles radiate part of their energy via synchrotron. We note that the assumption of power-law spectrum for non-relativistic electrons may not be realistic, however, this allows us to estimate the energy density in relativistic electrons as a function of time. In previous studies \citep{per14,cho16} it was assumed that all the electrons, in the post shock region, will be relativistic at any point of time. This introduces an artificial kink in their synchrotron radio light curve, and beyond this kink the light curve rises rather steeply.
\begin{figure*}
\centering
\includegraphics[width=8cm,angle=0]{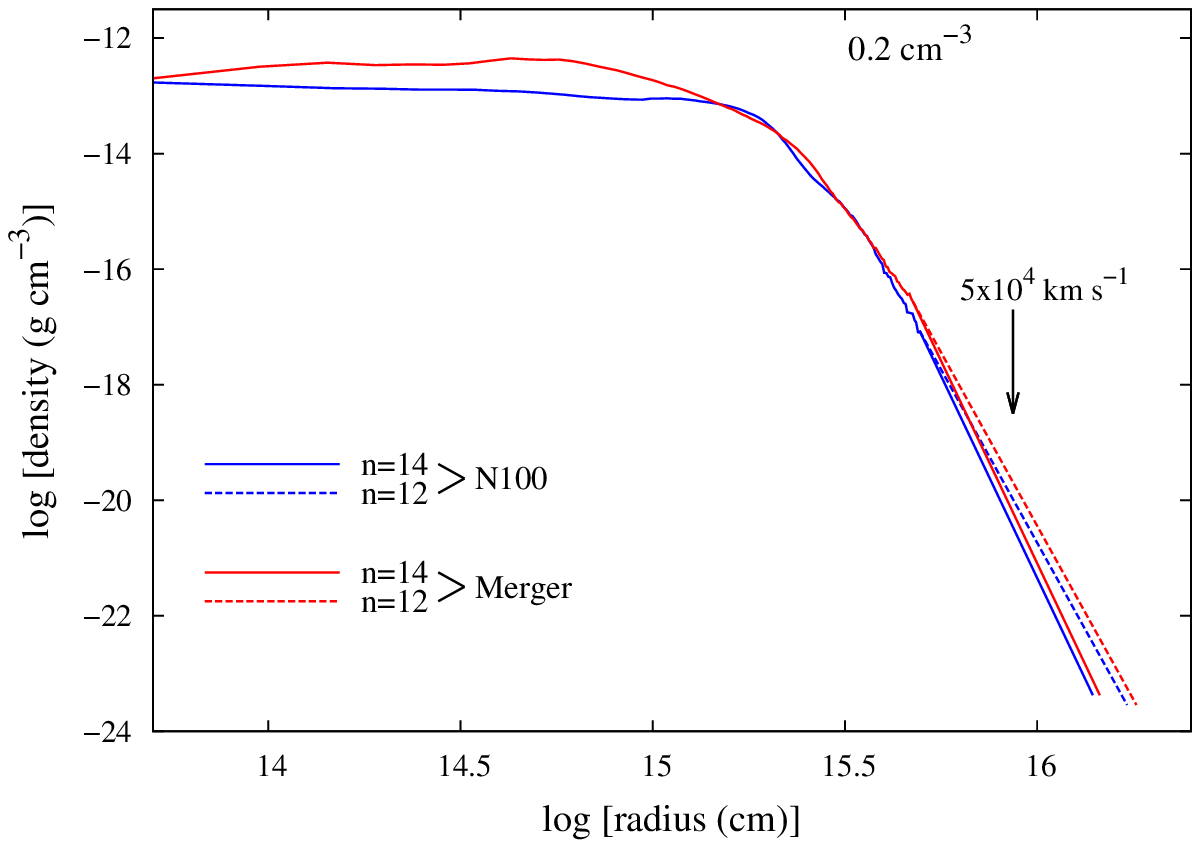}
\includegraphics[width=8cm,angle=0]{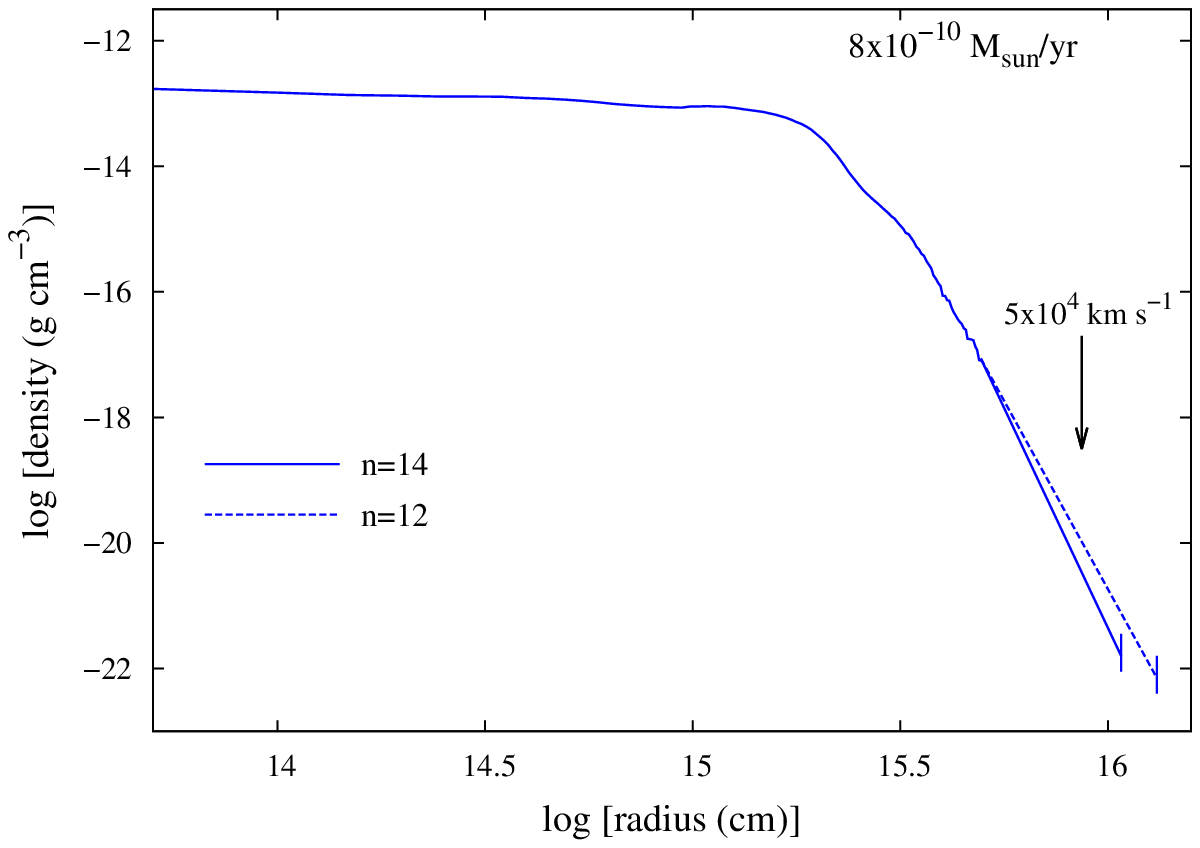}
\caption{Left panel: Unshocked ejecta structure as a function of radius for the N100 and merger models, at 20 days after explosion, when the ejecta are expanding into a constant density medium with $n_{\rm ISM}$ = 0.2 $\rm cm^{-3}$. Right panel: Same for the N100 model when the ejecta plough through a wind medium characterised by a mass loss rate of $\dot M = 8 \times 10^{-10}$\msunyr and wind speed of $100 \kms$. The arrow represents the radius beyond which the ejecta velocity is more than $5\EE4 \kms$. The structures are shown for two values of $n$ (12 and 14).   
}
\label{fig:ejstrucWindIsm}
\end{figure*}
Therefore, when the radio emissions predicted from such models are compared with observations, they suggest a rather low density medium around the progenitor system. As we consider the effect of shock deceleration in estimating $\epsilon_{\rm rel}$, and hence in the synchrotron light curve, our model predicts a more feasible density of the circumbinary medium where the shock was launched. \citet{cap14a} show that at non-relativistic shocks energetic particles could acquire 10-20$\%$ of the post shock energy. As particle acceleration is a complex mechanism and depends on the orientation of the background magnetic fields (see \citet{cap14a} for details) it is reasonable to consider the lower limit of acceleration efficiency. Therefore, for electrons we assume that 10$\%$ of the bulk kinetic energy is transferred to them, i.e. $\epsilon_{\rm e} = 0.1$, and kept it fixed at this value for our entire study.                

\par
A SN will be in the free expansion phase until the reverse shock remains in the outer power-law part of the ejecta. The unshocked ejecta profile as a function of radius at 20 days after the explosion, for both the N100 and merger models, are shown in figure \ref{fig:ejstrucWindIsm}. The left panel depicts the ejecta structure when the SN ploughs through a constant density medium with $n_{\rm ISM} = 0.2 \cm3$, and the right one shows that in a wind medium (for N100) characterised by $\dot M = 8 \times 10^{-10}~{\bigg(\frac{v_w}{100 ~\rm km~s^{-1}}\bigg)}^{-1}$\msunyr. As opposed to in \citet{per14}, where a thin-shell approximation was used for the shock structure, here we take into account the actual structure of the shocked gas from the similarity solutions \citep{che82a} for the various values of $s$ and $n$ investigated. 

The intensity of the synchrotron radiation, with synchrotron self absorption (SSA) as the main absorption mechanism, from a shell of thickness $\Delta r$ can be written as \citep{per14}
\begin{equation}
	I_{\nu}\left(h\right)=\frac{2kT_{\rm bright}} {c^2 f\left(\frac{{\nu}_{\rm peak}}{{\nu}_{\rm abs}}\right)}
	                 \frac{\nu^{5/2}} {\nu_{{\rm abs},0}^{1/2}}
	                 \left[
	                 1-{\rm exp}\left(-\xi_h \tau_{\nu_{0}}\right)
	                 \right],
\label{eq:Intensity}
\end{equation}
with 

\begin{equation}
	f (x) = x^{1/2} \left[1 - {\rm exp} \left( - x^{-\left(p+4\right)/2} \right) \right]
\label{eq:fx}
\end{equation}
\citep{bjo14}. Here $\xi_h  = \Delta s (h)/ (2 \Delta r)$ gives the normalised path length traversed by the radiation along the line of sight and $0 \leq h \leq 1$ takes into account emission from different parts of the shell. ${\nu}_{\rm abs}$ is the absorption frequency for which optical depth $\tau_{\nu_{\rm abs}} = 1$ and for $h = 0$ ${\nu}_{\rm abs} = {\nu}_{\rm abs,0}$ and $\tau_{\nu} = \tau_{\nu_{0}}$, respectively. $T_{\rm bright}$ and ${\nu}_{\rm peak}$ represent the brightness temperature and peak frequency of the radiation. $k$ is the Boltzmann constant. 

According to the models, SNe~2011fe and 2014J are in the optically thin regime when the radio observations are done (see $\S$ \ref{sec:data} for radio observations). Therefore, the luminosity, for $\tau_{\nu} < 1$, from a shell with outer radius $r_s$ can be written as following :      
\begin{equation}
	L_{\nu, {\rm thin}}=\frac{8 \pi^{2} kT_{\rm bright} \vartheta_{\nu} r_s^2} {c^2 f\left(\frac{{\nu}_{\rm peak}}{{\nu}_{\rm abs}}\right)}
	                 \nu_{{\rm abs},0}^{(p+3)/2} \nu^{-(p-1)/2},
\label{eq:Luminosity}
\end{equation}
with

\begin{equation}
	\nu_{{\rm abs},0}= \left(2 \Delta r ~ \varkappa(p) ~N_0 ~ B^{(p+2)/2}\right)^{2/(p+4)}.
\label{eq:nuabs}
\end{equation}
where $\vartheta_{\nu} = \frac{L_{\nu}}{L_{\nu,0}}$ with $L_{\nu,0}  =  4 \pi^2 r_{s}^{2} I_{\nu}(0)$. $B$ represents magnetic field strength and $\varkappa(p)$ is the SSA coefficient.  

\par
Observational evidence suggest that the brightness temperature, $T_{\rm bright}$, is $\sim$ $10^{11}$ K \citep{red94}. In our modelling we consider a value of $5 \times10^{10}$ K and keep it constant throughout the evolution, which is also what was assumed in \citet{per14}.  

\begin{figure}
\centering
\includegraphics[width=8.5cm,angle=0]{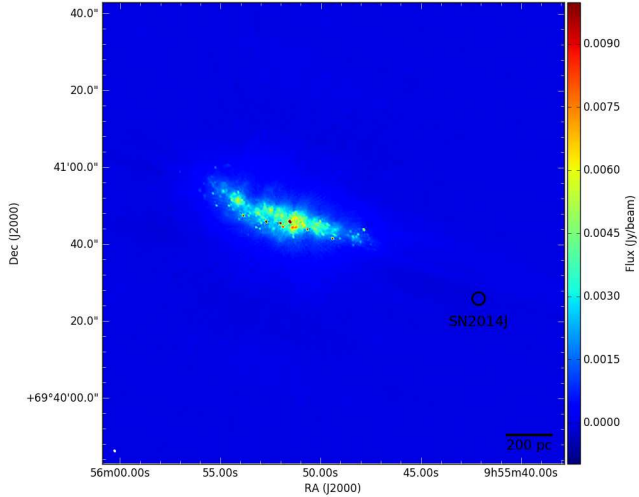}
\caption{JVLA radio image of the Type Ia SN~2014J and its host galaxy at a frequency of 3.0 GHz, obtained from our observations on August 2015. The total flux density is of 4.56 Jy, the peak is 47.8 mJy/beam, and the off-source rms of 7.4 $\mu$Jy/beam, attaining a dynamic range of 6530:1, which makes this the deepest image of M82 at this frequency ever. Note the large amount of compact sources that are detected in the field of view, most of them being young SNe and SNRs. The rms at the position of SN~2014J, marked with a circle, is 32 $\mu$Jy/beam, significantly larger than further away from the SN~2014J position and from the central regions of M82.
}
\label{fig:SN2014J}
\end{figure}

\section{Radio data}
\label{sec:data}

In this paper, we modeled the publicly available radio data for both SN~2011fe and SN~2014J, as well as radio observations obtained by us. Namely, for the modeling of SN~2014J, we used data published in \citet{per14}, complemented by recently obtained data with the Jansky Very Large Array (JVLA) at 3.0 GHz, and by the European VLBI Network (EVN) at 1.66 GHz. For the modelling of the radio emission from SN~2011fe, we used data published by \citet{cho12}, also complemented by recently obtained 3.0 GHz Jansky Very Large Array (JVLA) observations, and by 1.66 GHz European VLBI Network (EVN) observations. 

\subsection{New EVN observations} 

We observed SN~2014J on 28 February - 1st March 2015 (project EP092A; P.I. P\'erez-Torres), and SN~2011fe on 14 June 2015 (project EP092B; P.I. P\'erez-Torres), using  the EVN at 1.66 GHz, for 12 hr in each case (with an integrated on-source time of about 8.0 hr). Each time, we used a sustained data recording rate of 1024 Mbit s$^{-1}$, in dual-polarisation mode and with 2-bit sampling. Each frequency band was split into 8 intermediate subbands of 16 MHz bandwidth each, for a total synthesized bandwidth of 128~MHz. Each subband was in turn split into 32 spectral channels of 500~kHz bandwidth each.

The EVN observations of SN~2014J (EP092A) included the following 13-antenna
array: Effelsberg, Westerbork, Jodrell Bank, Onsala, Medicina, Torun, Urumqi,
Sheshan, Svetloe, Zelenchuk, Badary, Sardinia,  and the DSN antenna of Robledo
(Madrid), which joined the observations for about half of the total time,
although only recorded in LCP.  We note that Onsala could not take data due to
heavy winds and, most importantly, Effelsberg, our most sensitive antenna, could
not observe for 5 hr. Fortunately, we had additional large dishes for this
experiment, which allowed us to obtain a final r.m.s. value close to the
expected one (see below).   
We observed our target source, SN~2014J,
phase-referenced to the core of the nearby galaxy M81, known to be very compact
at VLBI scales, with a typical duty cycle of 5 minutes.  We used the strong QSO
1150+812 as fringe finder and bandpass calibrator. For our observations of
SN~2011fe (EP092B), the array included 10 antennas: Effelsberg, Westerbork,
Jodrell Bank, Onsala, Medicina, Torun, Svetloe, Zelenchuk, Badary, and Tianma.
Unfortunately, the data for Jodrell Bank and Onsala, were of very bad quality, so
we had to remove them completely, thus resulting in an eight-antenna array.
Further, the performance of Badary and, especially, Tianma (the second most
sensitive antenna), was rather poor. As a result, the attained r.m.s. was far
from the initially expected one, and significantly worse than attained for
EP092A (see below). 
We observed our target source, SN~2011fe, phase-referenced
to the strong, nearby source J1359+5544, known to be very compact at VLBI
scales, with a typical duty cycle of 5 minutes.  We used the strong source
3C309.1 as fringe finder and bandpass calibrator. 

All the data were correlated at the EVN MkIV data
processor of the Joint Institute for VLBI in Europe (JIVE, the
Netherlands), using an averaging time of 4~s for SN~2014J, and of 2~s for SN~2011fe.
 
We used \emph{AIPS}  for calibration,
data inspection, and flagging of our eEVN data, using standard procedures.  Those steps
included a-priori gain calibration (using the measured gains and
system temperatures of each antenna), parallactic angle correction and
correction for ionosphere effects.  We then aligned the visibility
phases in the different subbands, i.e., ``fringe-fitted'' the data, solved for the residual delays and 
delay rates, and interpolated the resulting gains into the scans of \sn.  We
then imaged a field of view of 3\arcsec$\times$3\arcsec centered at the position
given by \citet{smi14}, and applied standard imaging procedures using
\emph{AIPS}, without averaging the data neither in time, nor frequency,
to prevent time- and bandwidth smearing of the images. We used natural uv-weighting to maximize the signal to noise ratio in our final images, which was of 9.8 $\mu$Jy/beam and 34.3 $\mu$Jy/beam for SN~2014J and SN~2011fe, respectively.

\subsection{Jansky VLA observations} 
We also observed SN~2014J and SN~2011fe with the Jansky VLA on 15 - 16 August 2015 and 31 August 2015 (project 15A-076; P.I. P\'erez-Torres), respectively, while the VLA was in its most extended configuration.  Each target was observed for 2 hr, yielding an effective on-target time of $\sim$87 min. 3C286 was used as flux and bandpass calibrator. We centered our observations at the frequency of 3.0 GHz, covering the whole S-band (from 2.0 to 4.0 GHz), and recorded in full polarisaztion, using a dump time of 2 sec. The nearby, bright, point-like sources J0954+7435 (for SN~2014J), and J1419+5423 (for SN~2011fe) had been considered for phase-calibration. We used the \textit{Common Astronomy Software Applications} (\textit{CASA}, McMullin et al. 2007) package for data reduction purposes, which consisted of standard amplitude and phase calibration. Since our target sources were far from any bright point-like source, a natural weighting scheme was used to attain the best possible sensitivity in our images. In the case of SN~2011fe, we essentially reached the expected off-source r.m.s. value  (4.0 $\mu$Jy/beam). 
This was possible thanks to the fact that the VLA observations were carried
out with the array in its most extended configuration, i.e., with the smallest
possible synthesized beam at a given frequency.  In this way, most diffuse
emission was expected to be filtered out for M101, as was indeed the case.  
However, for SN~2014J, the
host galaxy, M82, a starburst galaxy, is extremely bright at low frequencies in the few central
kpc (see Fig.  \ref{fig:SN2014J}). In fact, the 3.0 GHz VLA emission is not limited by the thermal noise, since very far from the galaxy we actually reach values of about $7 \mu$Jy/beam,
very close to the nominal values. However, we are limited by the dynamic range
and the intrinsic background radio emission in the surroundings of SN~2014J, which is significantly higher, $\sim32.0 \mu$Jy/beam.

\par
Tables 1 and 2 show the summary of our new radio observations of SN~2011fe and SN~2014J, respectively, along with other existing radio data.

\begin{table}
\caption{Log of radio observations for SN~2011fe}
\scalebox{0.9}{
\begin{tabular}{lcclccr}
\tableline\tableline
Starting & Epoch & $t_{\rm int}$ & Array & $\nu$ & $S_\nu$ & {\bf L$_{\nu,23}$}   \cr
UTC        &   days &  hours          &           & GHz  & $\mu$Jy &  
 \cr
\tableline
%
 %
2011 Aug 25.77 &  2.1 &  $-$     & JVLA$^{\rm a}$        &   5.90   &   17.4   &   8.53           \cr         
2015 Jun 14 & 1390 & 8.0   & EVN$^{\rm b}$       &  1.66 &  103  & 50.5      \cr  
2015 Aug 31  & 1468 & 1.45   & JVLA$^{\rm b}$       &  3.00 &  12.0  & 5.88     \cr
\tableline
\end{tabular}
}
\tablecomments{The columns starting from left to right are as follows: starting observing time, UTC; mean 
observing epoch (in days since explosion, assumed to be on 2011 Aug 23.7);  integration time, in hr;
 facility; central frequency in GHz; 3$\sigma$ flux density upper limits, in $\mu$Jy; 
 the corresponding 3$\sigma$ spectral luminosity,  assuming a distance of 6.4 Mpc, in units of 
 $10^{23}$\,erg\,s$^{-1}$\,Hz$^{-1}$. Data from $^{\rm a}$\citet{cho12} and $^{\rm b}$This work.}
 \label{tab:RadioLog11fe}
\end{table}

\begin{table}
\caption{Log of radio observations for SN~2014J}
\scalebox{0.9}{
\begin{tabular}{lcclccr}
\tableline\tableline
Starting & Epoch & $t_{\rm int}$ & Array & $\nu$ & $S_\nu$ & {\bf L$_{\nu,23}$}   \cr
UTC        &   days &  hours          &           & GHz  & $\mu$Jy &  
 \cr
\tableline
%
 %
%
2014 Jan 23.2  &    8.2 &  $-$     & JVLA$^{\rm a}$        &   5.50   &   12.0   &   1.67           \cr           
2014 Jan 24.4  &    9.4 &  $-$   & JVLA$^{\rm a}$        &   22.0 &   24.0     &   3.31          \cr           
2014 Jan 28.8  &  13.8 &  13.6   & eMERLIN$^{\rm b}$ &  1.55 & 37.2      &   5.15       \cr
2014 Jan 29.5  &  14.5 &  14.0   & eMERLIN$^{\rm b}$ &  6.17 & 40.8     &     5.63    \cr
2014 Feb 4.0   &  20.0 &  11.0  & eEVN$^{\rm b}$       &  1.66 & 32.4  & 4.47   \cr
2014 Feb 19.1  &  35.1 &  10.0  & eEVN$^{\rm b}$      &  1.66 &  28.5  & 3.94    \cr
2014 Apr 11.2  &  86.2 &  $-$   & JVLA$^{\rm c}$       &  5.90 &  21.0  & 2.92   \cr
2014 Jun 12.0  &  148 &  $-$   & JVLA$^{\rm c}$       &  7.40 &  41.0  & 5.71   \cr
2015 Mar 1   &  410 &  8.0   & EVN$^{\rm d}$       &  1.66 &  29.4  & 4.10      \cr
2015 Aug 16  &  578  &  1.45  & JVLA$^{\rm d}$       &  3.00 &  96.0  & 13.4    \cr
\tableline
\end{tabular}
}
\tablecomments{The columns starting from left to right are as follows: starting observing time, UTC; mean 
observing epoch (in days since explosion, assumed to be on 2014 Jan 15.0);  integration time, in hr;
 facility; central frequency in GHz; 3$\sigma$ flux density upper limits, in $\mu$Jy; 
 the corresponding 3$\sigma$ spectral luminosity,  assuming a distance of 3.4 Mpc to M82, in units of 
 $10^{23}$\,erg\,s$^{-1}$\,Hz$^{-1}$. Data from $^{\rm a}$\citet{Chandler2014}, $^{\rm b}$\citet{per14}, $^{\rm c}$\citet{cho16} and 
 $^{\rm d}$This work.}
 \label{tab:RadioLog14J}
\end{table}

\begin{figure*}
\centering
\includegraphics[width=8.5cm,angle=0]{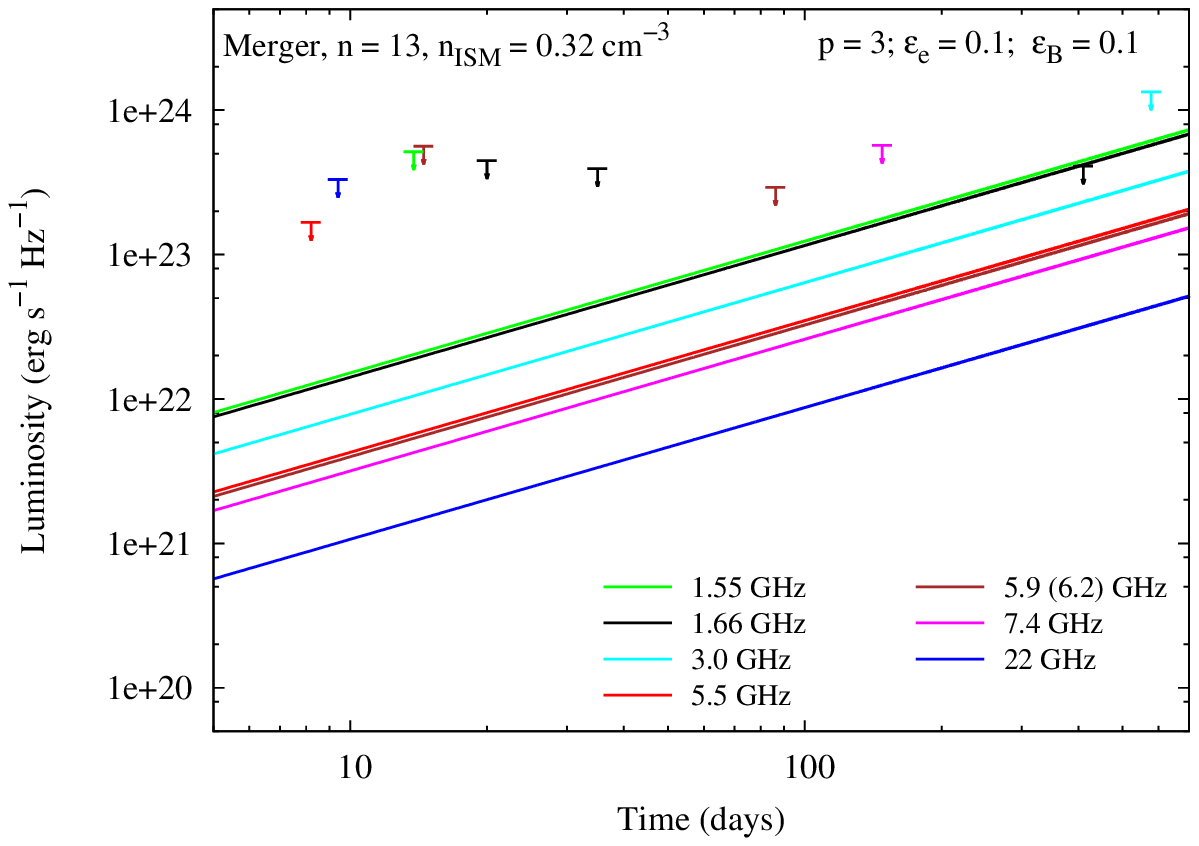}
\includegraphics[width=8.5cm,angle=0]{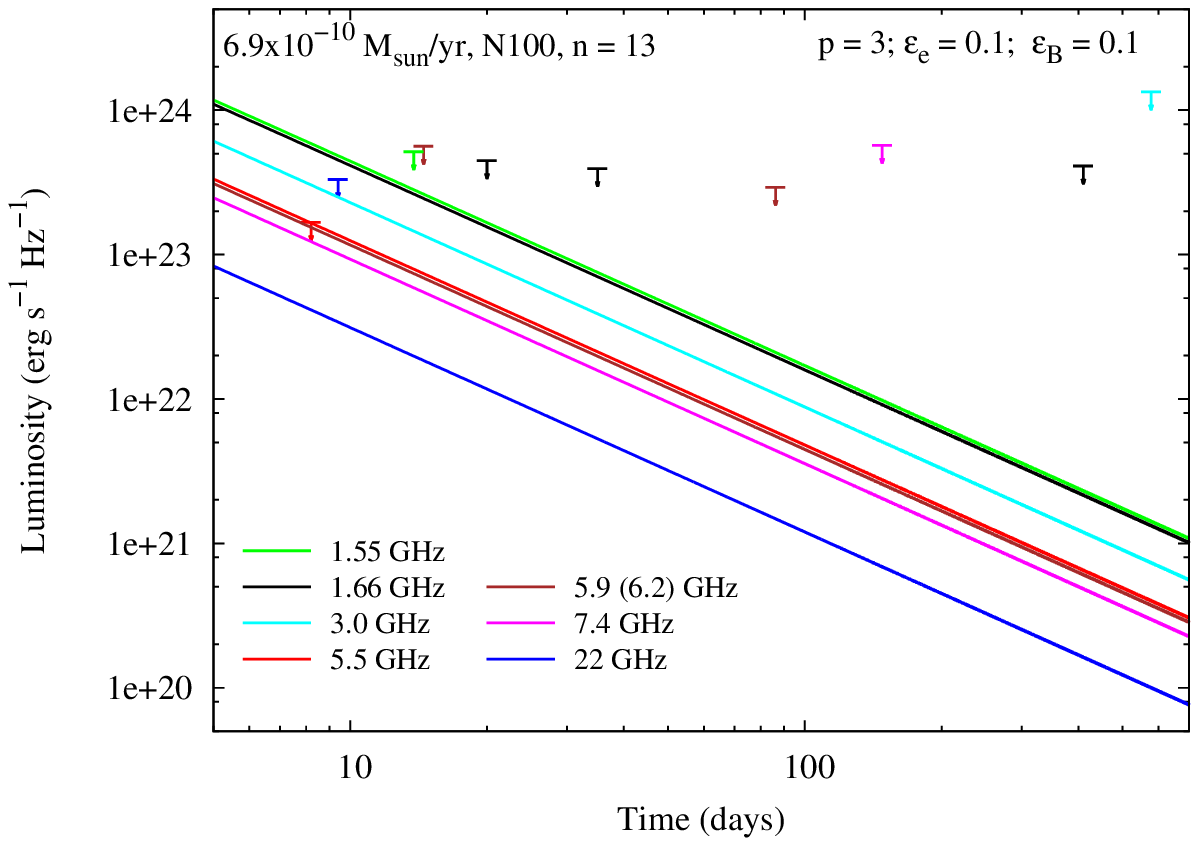}
\caption{Luminosity evolution as predicted by the merger model (left panel), shown with lines, for $n = 13$ and $\epsilon_{\rm B}$ = 0.1 in a constant density medium along with observational upper limits  obtained for SN~2014J (tabulated in table \ref{tab:RadioLog14J}). The upper limit on luminosity measured around 410 days after the explosion at 1.66 GHz constrains the particle density $n_{\rm ISM}$ to be $ < $ 0.32 $\rm cm^{-3}$ around the explosion site. The right panel shows the light curves from the N100 model, for the same parameters used in the merger model, when the SN is expanding in a wind medium. In this case the observation done around 8 days after the explosion at 5.5 GHz limits $\dot M < 6.9 \times 10^{-10}$$(v_w/100~{\rm km~ s}^{-1})^{-1}$~\msunyr{}.  
}
\label{fig:lcurves}
\end{figure*}

\vspace{1cm}
\section{Results}
\label{sec:results}
When the SN ploughs through a constant density medium the radio luminosity evolution, for an optically thin shell, can be written as following for $p = 3$ 
\begin{equation}
\begin{split}
    L_{\nu, {\rm thin}} \propto \bigg(\frac{n-3}{n} \bigg)^{3.86} ~ T_{\rm bright} ~ \epsilon_{\rm rel}^{1.71} ~ \epsilon_{\rm B}^{1.07}  ~  \nu^{-1} \\ 
    \left(n_{\rm ISM}\right)^{\frac{1.93n - 8.43}{n}} ~ t^{\frac{(2.86n - 25.3)}{n}},
\end{split}
\label{eq:Lumism1}
\end{equation}
and in the case of a wind stratified medium, $s=2$,
\begin{equation}
\begin{split}
    L_{\nu, {\rm thin}} \propto \bigg(\frac{n-3}{n-2} \bigg)^{3.86} ~ T_{\rm bright} ~ \epsilon_{\rm rel}^{1.71} ~ \epsilon_{\rm B}^{1.07}  ~  \nu^{-1} \\ 
    \left(\dot M / v_w\right)^{\frac{1.93n - 8.43}{n - 2}} ~ t^{-\frac{(n + 2.57)}{n-2}}.
\end{split}
\label{eq:Lumwind1}
\end{equation}
Here $\epsilon_{\rm B} = \frac{u_{\rm B}}{u_{\rm th}}$,  where $u_{\rm B} = B^2/8 \pi$ is the amount of post shock energy density in magnetic fields\footnote{The exponent of $(\dot M / v_w)$ in equation 6 of \citet{per14} should be 1.58 rather than 1.27 mentioned in that paper. The erroneous exponent does not affect the results of that paper.}.
\par 
We consider here two cases; 1) equipartition of energy between electric and magnetic fields, i.e., when $\epsilon_{\rm e}$ = $\epsilon_{\rm B}$ = 0.1. This implies that each field carries 10 $\%$ of the bulk shock energy. 2) $\epsilon_{\rm e}$ = 0.1 and $\epsilon_{\rm B}$ = 0.01, i.e. when 10$\%$ and 1$\%$ of post shock energy are in electric and magnetic fields, respectively. Therefore, in the second situation the amplification of magnetic field, in the post shock region, is less than that in the former one. 
  In the paper when we mention a low amplification efficiency of magnetic fields/ the magnetic field amplification is less than that presumed for energy equipartition/ $\epsilon_{\rm B}$ = 0.01 we refer to the second case. The first case where energy equipartition is assumed is cited as $\epsilon_{\rm B}$ = 0.1 throughout the paper. 

\par 
For $\epsilon_{\rm B}$ = 0.1 and $n = 13$ the light curves predicted by the merger model are shown in the left panel of fig.\ref{fig:lcurves} for SN~2014J. As can be seen, the density of the CSM is constrained by the upper limit at 1.66 GHz measured around 400 days after the explosion. This enforces the density of the surroundings to be $< 0.32$ $\rm cm^{-3}$ for a constant density medium. For the same parameters, the radio light curves calculated using the N100 model in a wind medium is depicted in the right panel of fig.\ref{fig:lcurves}. In this case, the mass loss rate of the progenitor system is restricted by the observational upper limit obtained at 5.5 GHz, around 8 days after the explosion of SN 2014J \citep{per14}. According to our model this implies $\dot M < 6.9 \times 10^{-10}$$(v_w/100~{\rm km~ s}^{-1})^{-1}$~\msunyr{}(right panel of fig.\ref{fig:lcurves}).

\par
As for SN 2014J, SN~2011fe was also thoroughly searched for in the radio just after the explosion \citep{horesh12, cho12,cho16}. Among the early observations, we found that the upper limit on the luminosity at 5.9 GHz, measured around 2 days after the explosion, \citep{cho12} (tabulated in table \ref{tab:RadioLog11fe}) constrains the maximum possible mass loss rate from the system. This was also noted by \citet{cho12} and \citet{per14}. In case of a constant density medium the radio luminosity
increases with time (eqn.\ref{eq:Lumism1}), when $ 12 \leq n \leq 14$,  in contrast to the wind medium for which the luminosity decreases with time (eqn.\ref{eq:Lumwind1}) (see the trends of the light curves of SN~2014J displayed in fig.\ref{fig:lcurves} for wind and constant density medium when $n = 13$). It is therefore found that for a constant density circumbinary medium the radio upper limit measured around 4 years after the explosion at 3 GHz limits $n_{\rm ISM}$. Table \ref{tab:RadioLog11fe} contains the details of these observations which are relevant to derive the upper limits on the CSM density, together with the hitherto unpublished limits measured at 1390 days at 1.66 GHz.      

\par
\begin{figure*}
\centering
\includegraphics[width=8.5cm,angle=0]{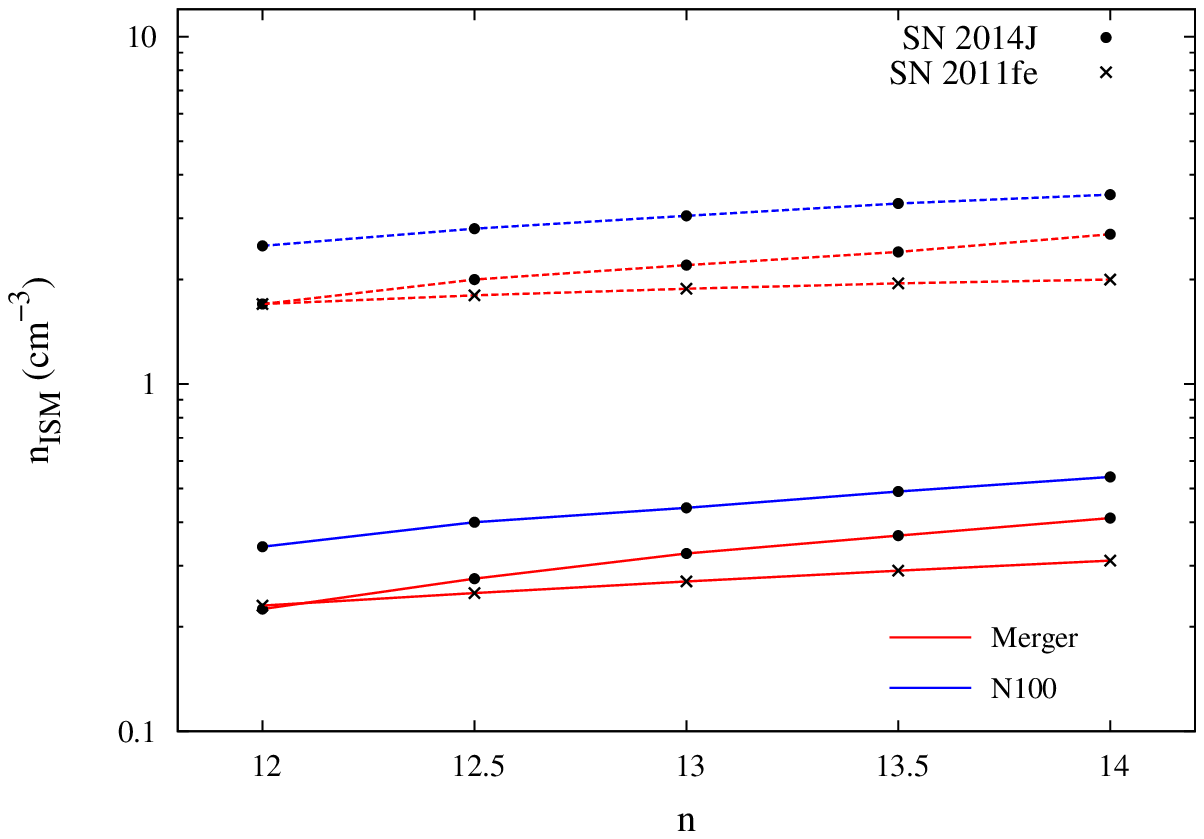}
\includegraphics[width=8.5cm,angle=0]{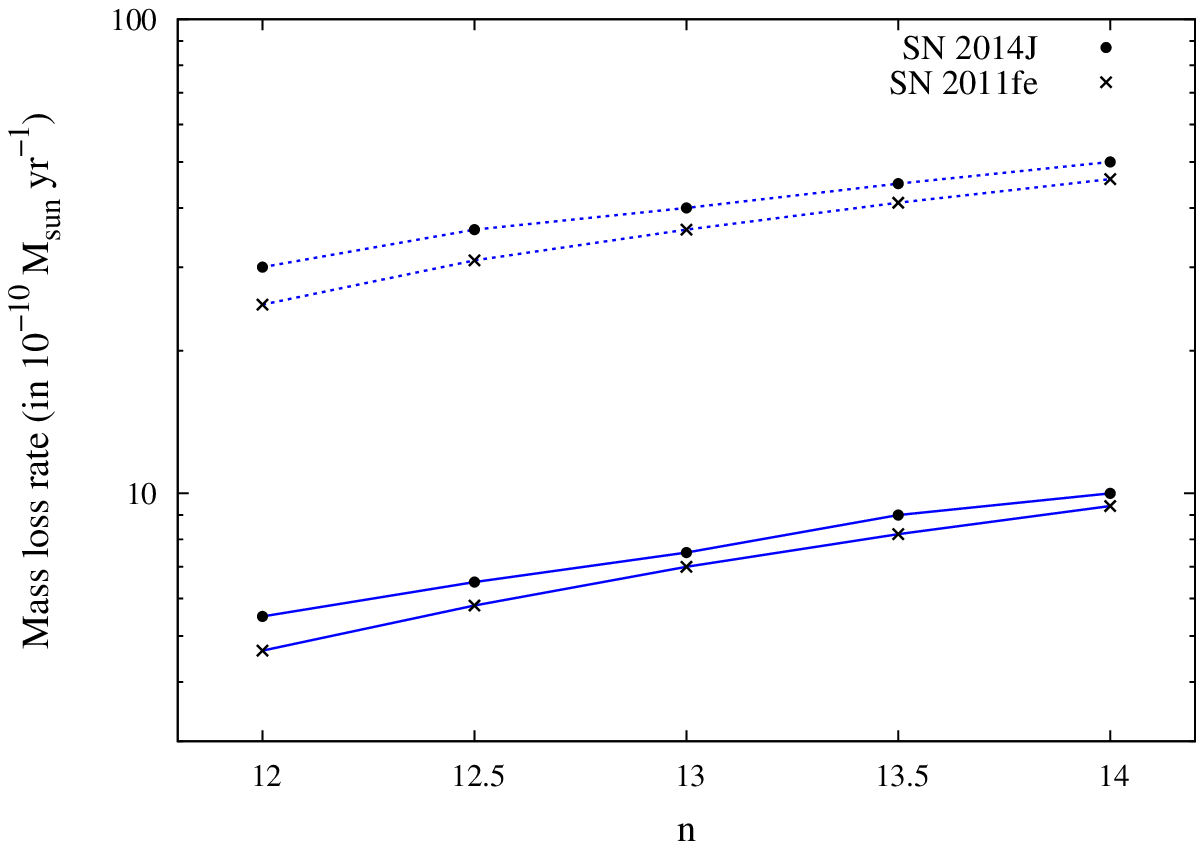}
\caption {Predicted upper limits on $n_{\rm ISM}$ (left panel) and $\dot M$ with $v_w =$ 100 $\rm km~s^{-1}$ (right panel) by the N100 (blue) and merger (red) models as a function of $n$. Any line with filled circles (or crosses) represents the dependence for SN~2014J (SN~2011fe).The solid lines are for $\epsilon_{\rm B} = \epsilon_{\rm e}$ = 0.1, i.e., when equipartition is assumed. The dashed lines show the effect of a low magnification efficiency of the magnetic field, i.e., $\epsilon_{\rm B}$ = 0.01, keeping $\epsilon_{\rm e}$ fixed at 0.1.
}
\label{fig:const_den_mloss_effect}
\end{figure*}

In fig.~\ref{fig:const_den_mloss_effect}, we show the dependence of the CSM density  on $n$ using N100 (blue) and merger model (red). The solid lines with crosses (filled circles) represent the upper limits on $n_{\rm ISM}$ or $\dot M$ for $v_w =$ 100 $\rm km~s^{-1}$ around SN~2011fe (SN~2014J) when $\epsilon_{\rm B} = 0.1$. The dashed lines show the effect of a low magnetic field energy density, $\epsilon_{\rm B}$ =  0.01.
As is clear from the figure, there is little variation on the upper limits of $n_{\rm ISM}$ when $n$ varies in range 12-14. In case of a wind medium, $\mdot /v_w$, calculated from N100, has a rather strong dependence on $n$. The upper limits on $n_{\rm ISM}$, from both models, are  $\sim 0.35~{\rm cm}^{-3}$ when equipartition of energy between electric and magnetic field is considered, i.e., $\epsilon_{\rm B} = \epsilon_{\rm e}$ = 0.1. For a low magnetic field energy density ($\epsilon_{\rm B} = 0.01$), which implies a low amplification efficiency for magnetic fields in the shocked region, the upper limits are $\sim 2~{\rm cm}^{-3}$. The upper limits on mass loss rate predicted by the N100 model, for SNe~2011fe and 2014J, are $ < 10^{-9}$ \msunyr and $< \sim 4 \times 10^{-9}$ \msunyr for $\epsilon_{\rm B} = $ 0.1 and 0.01, respectively, assuming $v_w = 100 \kms$.

Several authors have calculated the upper limits on $\mdot$ and $n_{\rm ISM}$, for both SNe 2011fe and SN 2014J, considering $\epsilon_{\rm e} = \epsilon_{\rm B}$ = 0.1 and $v_w$ = $100 \kms$. From a very early non-detection in radio, \citet{horesh12} constrained $\mdot$ of SN 2011fe to be $< 10^{-8} \msunyr$ assuming $v_s$ = $4 \times 10^{4}$ $\kms$. However, just after the explosion, the shock velocity could be much higher than $4 \times 10^{4}$ $\kms$ \citep{ber02}. \citet{cho12} estimated $\mdot < 6\times 10^{-10}~\msunyr$ ($n_{\rm ISM} < $ 6 cm$^{-3}$) using their upper limits on radio luminosity, which were deeper than those reported in \citet{horesh12}. For SN 2014J, \citet{per14} found $\mdot < 7\times 10^{-10}\msunyr$ ($n_{\rm ISM} < $ 1.3 cm$^{-3}$) from their radio analysis.  From X-ray observations, where $\epsilon_{\rm B}$ plays no role, \citet{mar12,Margutti2014} obtained $\mdot  < 2\EE{-9}$~\msunyr{} and $< 1.2\EE{-9}$~\msunyr{} ($n_{\rm ISM} < $ 166 cm$^{-3}$ and $<$ 3.5 cm$^{-3}$) for SNe~2011fe and 2014J, respectively. In our present analysis we used an ejecta structure and a radio emission model similar to those presented in \citet{cho12} and \citet{per14}. We estimate $\mdot < \sim 7\times 10^{-10}~ \msunyr$ ($n_{\rm ISM} < \sim $ 0.35 cm$^{-3}$) (see fig.~\ref{fig:const_den_mloss_effect}) for both SNe, when $\epsilon_{\rm e} = \epsilon_{\rm B}$ = 0.1 and $v_w = $ 100 $\kms$. 
This shows that our upper limits on $\mdot$ are consistent with the previous findings by \citet{cho12,per14}. However, our upper limits on $n_{\rm ISM}$ is almost an order of magnitude smaller than those reported from previous radio analyses. The reason is that to derive the limits \citet{cho12,per14} used the upper limits on radio luminosity measured around 20 (for SN~2011fe) and 35 (for SN~2014J) days after the explosion, which was the best possible option at that time. In our present study we obtained the upper limits  from radio observations carried out around 4 and 1.5 years after the explosion of SNe 2011fe and 2014J, respectively. Therefore, this study provides a deeper and more confident limit compared to the previous ones. The limits on $\mdot$ reported from X-ray analyses are midway between our predictions from $\epsilon_{\rm B}$ = 0.1 and 0.01.

\section{Discussion}
\label{sec:discussion}
SN~2006X, a normal SN~Ia, showed notable presence of circumstellar material in the form of time evolution of a blue shifted absorption feature in the Na~I~D~line \citep{pat07}. This SN was radio silent despite prompt attempt was made to observe the SN in radio wavelengths \citep{stock06}. We note that the optical and radio observations were not conducted simultaneously.  
However, given the fairly large distance to SN~2006X ($\sim 16$ Mpc\footnote{From the NASA/IPAC Extragalactic Database (NED)}), it could be difficult to detect radio emission from it if magnetic field amplification is low in SN shocks. 

\par
No variation in Na~I~D was reported for SN~2011fe \citep{pat11} or for SN~2014J \citep{goo14}. The lack of narrow line absorption could be due to a viewing angle effect, allowing for incomplete shells to possibly exist around those SNe. How frequently one could expect to detect such shells cannot be predicted in advance, unless proper hydrodynamical modeling of the progenitor evolution and shock CSM-shell interaction are done. With our observations we are probing a region with radius of few $\times 10^{17}$ cm around both the SNe. The radio silence feature of these SNe could be due to no shells existing around them within the above mentioned radius. 
Or it may be due to low amplification efficiency of magnetic fields in the shocks. We try to explore the second possibility with our modeling of the radio observations, tabulated in tables \ref{tab:RadioLog11fe} and \ref{tab:RadioLog14J} for  SNe~2011fe and 2014J, respectively. 

As discussed in $\S$ \ref{sec:results} the evolution of radio luminosities (eqns. \ref{eq:Lumism1} and \ref{eq:Lumwind1}, and fig.\ref{fig:const_den_mloss_effect}) depend on the nature of circumbinary medium which carries the signature of the progenitor system. The progenitors of both SNe~2011fe and 2014J are not clear. However, observational evidences have narrowed down the possible formation channels for both of them. 
From pre-explosion HST images \citet{kelly14} have excluded a bright red giant (RG) companion and a symbiotic recurrent nova system, having similar luminosities of RS Oph, as a progenitor system for SN~2014J. In case of SN~2011fe, \citet{li11} excluded a RG/helium (He) star/main sequence (MS) with mass more than $3.5~\msun$ as a companion. From UV observations \citet{bro12} precluded a solar mass MS companion, with a radius of 0.1~\rsun~\citep{bloom12}, although their model suffers from a viewing angle effect.
Combining these findings with non-detections of binary companion material in optical nebular spectra of the two SNe, \citet{lun15} conclude that hydrogen-rich main-sequence donor systems, with a separation less than $\sim 7$ \rsun between the binary systems, could be ruled out for SN~2011fe, while there is still some room for this for SN~2014J. In addition, He star donors are constrained to those with large separation (more than $\sim$ 1 \rsun) between them and the WD. 

\par
These studies favour a DD formation channel for both the SNe. However, a few SD channels are still allowed. For these formation channels we will now examine what upper limits on $n_{\rm ISM}$ and $\mdot$ our models, with $\epsilon_{\rm B}$ = 0.1 and 0.01, predict and what is expected for that kind of progenitor.

\subsection{DD scenario}
\label{section:DD}
The merger of two WDs predicts a low gas density, which is characteristic of the interstellar medium, around the SN site. The H~I column density, $N_{\rm HI}$, around SN~2011fe is $3 \times 10^{20}$ $\rm cm^{-2}$ as estimated by \citet{cho12}. With a path length ($l$) of $\sim 100$ pc and mean atomic weight $\mu \simeq 1.24$ in M101 \citep{stoll11,mazzali14}, the particle density around the location of SN~2011fe is $\mu  N_{\rm HI}/l \sim 1 $ $\rm cm^{-3}$. 
From our modeling with the merger model, the null detection gives an upper limit on the particle density $n_{\rm ISM} \sim $ 1~cm$^{-3}$ (left panel of fig.\ref{fig:const_den_mloss_effect}) when $u_B < u_e$ with $\epsilon_{\rm B}$ = 0.01. For $\epsilon_{\rm B} = 0.1$ the upper limit on the particle density is $\sim 0.25\rm ~cm^{-3}$ . 

\par
SN 2014J occurred in M82. \citet{ritchey15} estimated $N_{\rm HI}$ in the direction of SN~2014J from 21~cm line observations. They obtained $N_{\rm HI} = 2.9 \times 10^{21}$ $\rm cm^{-2}$ for the gas around the SN site and compared it with $N_{\rm HI}$ computed from the equivalent width, contributed by M82 along the supernova line-of-sight, of the blended $\lambda$5780.5 diffuse interstellar band (DIB). From the DIB they derived $N_{\rm HI} = 2.24 \times 10^{21}$ $\rm cm^{-2}$ which is close to that derived from the 21~cm emission.
The projected distance of SN~2014J from the center of M82 is $\simeq 913$ pc for a distance to M82 of 3.4 Mpc (see Fig.\ref{fig:SN2014J}). Combining this with the H~I column density map shown in the upper left panel of Fig.7 of \citet{leroy15}, we estimate $l \sim 1$ kpc for the atomic hydrogen gas around SN~2014J in M82. Here we assume that the SN is located at the middle of the disk. With $N_{\rm HI} \simeq 2.9 \times 10^{21}$ $\rm cm^{-2}$, along with slightly sub-solar metallicity of M82 \citep{ori04}, this implies a particle density of $\sim$ 1 $\rm ~cm^{-3}$ around the SN location. The left panel of fig.\ref{fig:const_den_mloss_effect} displays that the radio non-detection can be well accounted for with the above expected ISM density by assuming $\epsilon_{\rm B} = 0.01$.
The effect of a higher value of $\epsilon_{\rm B}$ (= 0.1) on $n_{\rm ISM}$ is similar to that for SN~2011fe.      
    
\subsection{SD Channel}
\label{section:SD}
Considering the observational constraints on the progenitor systems of both the SNe, a He star with large separation ($\gsim$ 1 \rsun) could be suggested as a binary companion to the WD. For SN~2014J a hydrogen rich MS may also be possible, although the separation to the WD is likely to have been $\gsim 5 \rsun$ \citep{lun15}. In these cases it is expected that before explosion mass has been transferred from the secondary to the WD via Roche Lobe overflow.
As mentioned in $\S$\ref{sec:results} and shown in the right panel of fig.\ref{fig:const_den_mloss_effect}, for SNe~2011fe and 2014J, the upper 
limit on $\mdot$ is $ < 10^{-9}$~$(v_w/100~{\rm km~ s}^{-1})^{-1}$~\msunyr{} for $\epsilon_{\rm B} = $ 0.1. 
This implies that before the explosion the WD accreted matter with an efficiency more than 99$\%$.
However, after studying radio observations of 27 SNe~Ia, \citet{panagia06} suggest that in case of Roche Lobe overflow the acceretion efficiency could be maximum 80$\%$. Therefore, for this kind of companion, a rather dense medium is expected with a higher value of $\mdot /v_w$. From our model with $\epsilon_{\rm B} = $ 0.01 we estimate $\mdot < \sim 4 \times 10^{-9}$~$(v_w/100~{\rm km~ s}^{-1})^{-1}$~\msunyr{} for both the SNe. This implies a less restricted mass accretion efficiency of the progenitor WD in comparison to the case with $\epsilon_{\rm B} = $ 0.1.
  
Within the SD scenario two other possibilities exist, namely the spin up down model and nova recurrence. We discuss them next.

\subsubsection{Spin up/down Model}
\label{section:SDspin}
As discussed in \citet{hachisu12a, hachisu12b}, rigidly rotating WDs within the mass range $1.38~\msun < M_{\rm WD} \lsim 1.5~\msun $, and with a MS or RG companion, can result in normal SNe~Ia. For such WDs the spin down time, determined by the magnetodipole radiation, can be $>10^{9}$ yrs when they have low magnetic fields and explode with low angular momentum \citep{ilkov12}. Therefore, the MS or the RG companion may evolve into a helium WD (in case of RG, a C/O WD is also possible to form as the evolution timescale for RG to form a C/O degenerate core is $< 10^{8}$ yrs). This spin down time is enough to allow the CSM, created by the mass loss from the progenitor system, to diffuse into the ISM. Therefore, the primary WD explodes in an ISM like environment. In such a system, the companion could also have negligible imprint on optical nebular spectra \citep{jus11}, even if it did not yet evolve into a WD \citep{lun15}.

We examine the N100 model for SN~2014J, in a constant density medium, considering  $\epsilon_{\rm B}$ = 0.1 and 0.01. 
N100 is an explosion model for a non-rotating WD in contrast to the spin up/down model which takes into account WD rotation. We, however, expect rotating WDs in the above mentioned mass range to have an overall similar ejecta structure to that of N100 as both of them could account for normal type Ia. 
Therefore, it can be expected that the prediction from N100 could be close to that from a spin up/down model.
From fig.\ref{fig:const_den_mloss_effect} (left panel) the upper limit on the surrounding density is $\sim 3  \cm3$ for $\epsilon_{\rm B}$ = 0.01 , which is similar to what could be expected around SN~2014J (see $\S$ \ref{section:DD}). For $\epsilon_{\rm B}$ = 0.1, the limit is $\sim 0.4 \cm3$. 

\par
A similar spin up/down model is applicable to SN~2011fe \citep{hachisu12b}. For $\epsilon_{\rm B}$ = 0.1, and using our radio observations from 1468 days, the upper limit on $n_{\rm ISM}$ is $\sim 0.3 \cm3$. However, from SN evolution with $\epsilon_{\rm B}$ = 0.01, we note that, in our models, around 3 years after the explosion the reverse shock starts to move inside the steep outer density profile of the ejecta.
This implies that the  self-similar solution, which is applicable as long as the SN is in the free expansion phase, is no longer valid. The SN then evolves into the Sedov-Taylor (ST) phase. The duration of free expansion can be written as $T_{FE} = \Lambda^\frac{1}{(3-s)} ~ \zeta_2^\frac{n-s}{3-s} ~ v_{o,ej}^\frac{s}{3-s}$ where $\Lambda = \frac{4 \pi}{\theta} ~ \bigg( \frac{3-s}{n-3} \bigg) ~ \frac{\rho_{o,ej}}{\beta} ~ \frac{\zeta_2^{3-n}}{\zeta_1^{3-s}} ~ r_{o,w}^{2-s}$. Here $\beta = \mdot/v_w$ ($n_{\rm ISM}$) for a wind (constant density) medium. $\theta$ is the ratio between swept up ejecta and CS mass. $\rho_{o,ej}$ and $v_{o,ej}$ represent the density and velocity of the extreme outer part of the inner ejecta. $\zeta_1 = r_s/r_c$ and $\zeta_2 = r_{rv}/r_c$ with $r_c$ and $r_{rv}$ being the contact discontinuity and reverse shock radii. $r_{o,w}$ represents a reference radius and this is  $10^{15}$ cm in our case. It is found that for a wind medium $T_{FE}$ is $> 100$ years for the range of mass loss rates we consider here. However, in case of a constant density medium, the ST phase starts much earlier. Using the merger model we estimate that $T_{FE}$ is around 8 and 5 years (10 and 5 years) for SN 2014J (SN 2011fe) when $\epsilon_{\rm B}$ = 0.1 and 0.01, respectively. In case of the N100 $T_{FE}$ is comparably shorter for SN 2014J (SN 2011fe), around 5 (6) years for $\epsilon_{\rm B}$ = 0.1. When $\epsilon_{\rm B}$ = 0.01 the ST phase starts $\sim 3$ years after the explosion of both the SNe with $n_{\rm ISM} \sim$ 1~cm$^{-3}$. For SN 2011fe, $n_{\rm ISM}$ is constrained by the radio luminosity measured at 1468 days, during which the SN is in the ST phase for $\epsilon_{\rm B}$ = 0.01. Modeling synchrotron emission from this phase is beyond the scope of this paper. This, therefore, limits our conclusions for the spin up/down scenario for SN~2011fe when $\epsilon_{\rm B}$ = 0.01.

\subsubsection{Nova recurrence}
\label{section:SDnova}
A near-Chandrasekhar mass WD, accreting at a rate between $10^{-7} - 10^{-12}$ \msunyr, experiences nova eruptions with a recurrent time, $t_{rec}$, of few years to hundred thousand years \citep{yaron05}. The maximum expansion velocity of the ejected shells is in the range $1000 - 5000 \kms$. This nova shell sweeps up the CSM material and a cavity is formed around the progenitor system. If $t_{rec}$ is long, the cavity is refilled by the new stellar wind. In this case one would expect eqn.\ref{eq:Lumwind1} to govern the radio luminosity evolution. For short $t_{rec}$, the cavity will be partially refilled. Therefore, the SN ejecta first encounter a wind medium, then move into a cavity, and finally interact with a nova shell. There could be multiple nova shells present in the circumbinary medium. The evolution of radio synchrotron spectra in this scenario can only be predicted by hydrodynamical modeling, which takes into account the interaction of SN ejecta with different kinds of ambient media, formed at different times of the progenitor evolution.

Recently, \citet{harris16} have done hydrodynamical modeling of a SN shock interacting with CSM shells, to predict the synchrotron emission from such a situation. The authors consider vacuum between the exploding WD and the interacting CSM shells. In their model the radio light curve has a plateau when the SN ejecta with velocity $2\times10^4 \kms$ interact with two nova shells. This feature is distinctly different from that of the supernova interacting with a single nova shell, for which a peak in the light curve is predicted. \citet{harris16} show that when the ejecta velocity is increased to $3\times10^4 \kms$, the peak radio luminosity decreases. Along with shock velocity, the density and the thickness of the shell, which are both poorly constrained, play important roles in predicting the synchrotron emission from the SN. 
As several less constrained physical parameters are involved here, it is from our models difficult to predict the expected radio luminosity for this kind of system. 
However, from very early X-ray non-detection \citet{Margutti2014} have precluded a recurrent nova system, with recurrence time of 100 yrs \citep{dim14}, for SN 2014J. This is also true for SN 2011fe as this SN was not seen in early X-ray observations as well \citep{mar12}.  
Moreover, given the constraints on possible SD binary companions \citep[cf.][]{lun15} a nova-shell like scenario may not be applicable for these SNe, as is also indicated by the absence of Na~I~D variations \citep{pat11,goo14}. 
The only absorption-line feature detected for these SNe is in K~I~$\lambda 7665$ \citep{gra15,mae16}, but the absorbing gas is at distances of at least $10^{19}$~cm from the supernova, and could be of interstellar origin \citep{mae16}.

\section{Conclusion}
\label{sec:con}
We have observed the SNe~Ia 2011fe and 2014J in the radio at late epochs, up to 1468 and 578 days post explosion, respectively. We do not detect the SNe, and model the radio non-detections assuming synchrotron emission from shock accelerated relativistic electrons to be the origin of the radiation. We consider that all the electrons in the post shock region are accelerated and their energy distribution follow a power-law profile with $p = 3$. It is assumed that 10$\%$ of bulk shock energy is converted to  electrons, i.e. $\epsilon_{\rm e}$ = 0.1. After a certain time, as the shock slows down, a part of the electron population will have kinetic energy less than the rest mass energy of electron. We calculate the fraction of energy that goes into these non-relativistic electrons and estimate the synchrotron flux from the energy available to relativistic electrons. This allows us not to underestimate the density of the ambient medium. 

\par
Both SNe~2011fe and 2014J are normal type Ia. We have therefore used the inner density structure of the N100 model (for the SD channel) and a violent merger model (for the DD channel), which could both account for normal SNe~Ia, and added a power law density profile, with index in the range 12-14, for the outer part of the ejecta. Considering that the interaction of SN ejecta with the CSM follows  a self-similar structure, we compare the radio light curves predicted by our models with observational upper limits. This gives us the upper limits on the 
ambient medium density, i.e., $n_{\rm ISM}$ for a constant density medium and $\mdot / v_w$ for a wind medium. Since synchrotron power is proportional to the magnetic field energy density, the amplification of this field, in the shock, plays an important role in determining the synchrotron luminosity, and hence on the prediction of the ambient medium density.

\par 
 The circumbinary medium is shaped by the mass loss history of the progenitor system. Although the kind of  binary systems that result in these two SNe are not clear, studies based on observational evidences \citep{li11,bro12,bloom12,kelly14,lun15} have given indications about the formation channels of these two SNe.
Among these, the DD scenario is one of the possibilities. From the H~I column density observed around both the SNe, $n_{\rm ISM}$ could be expected to be $\sim 1 \cm3$. To estimate $n_{\rm ISM}$ from the H I surface density, we assume a path length of 100 pc and 1 kpc for the atomic hydrogen gas in M101 and M82 around SNe 2011fe and 2014J, respectively. From our modeling we found that (left panel of fig. \ref{fig:const_den_mloss_effect}) the radio non-detections post one year can be explained with $n_{\rm ISM} \sim$ 1 $\cm3$ when the magnetic field amplification is low ($\epsilon_{\rm B}$ = 0.01) compared to what is expected for equipartition of energy (i.e., $\epsilon_{\rm B}$ = $\epsilon_{\rm e}$ = 0.1). 

As discussed in $\S$ \ref{section:SD}, a main sequence companion may be possible for SN~2014J in the SD scenario, whereas this is less likely for SN~2011fe. A He star companion is easier to accommodate for both SNe. In this channel, we found that the upper limits on $\mdot / v_w$ predicted by our model, considering a low value of $\epsilon_{\rm B}$, could allow for significant mass loss (see right panel of fig. \ref{fig:const_den_mloss_effect}). For $\epsilon_{\rm B} = 0.1$ and $0.01$, the upper limit on $\mdot$ for both the SNe are $ \sim 7 \times 10^{-10}$ and  $ \sim 4 \times 10^{-9}$  ~\msunyr when $v_w = 100 ~{\rm km~ s}^{-1}$. This suggests that in the former case the mass accretion efficiency of the WD needs to be more than 99$\%$, whereas the later one puts a less constraining limit on accretion efficiency. From a radio study of 27 Type Ia SNe, \citet{panagia06} found a maximum accreting efficiency of 80$\%$ for this kind of progenitor system.
   
Another possibility within the SD channel is recurrent novae. In this case, the wind may be swept up in shell-like structures prior to the SN explosion. Our models cannot predict the radio emission for this scenario as the SN could interact  with different kinds of ambient media depending on its evolution. To predict the synchrotron luminosity from this kind of system a proper hydrodynamical modeling is required. However, considering the absence of Na~I~D variation, early non-detections in X-rays, as well as no indications of hydrogen and metals from an SD companion in optical nebular spectra of the two SNe, recurrent novae may be less likely progenitor systems for these SNe. 
    
A particularly interesting case for the SD scenario is instead the spin up/down model which suggests a density around the progenitor system similar to the ISM density. We have tried to explore this channel, for SN~2014J, by considering an SN, having a density profile of the N100 model for the inner part of the ejecta, expanding into a constant density medium. The upper limits on $n_{\rm ISM}$ predicted by this model (left panel of fig. \ref{fig:const_den_mloss_effect}) is $\sim$ 3 $\cm3$ when $u_B < u_e$ with $\epsilon_{\rm B}$ = 0.01.

\par
Modeling of radio and X-ray emissions from core collapse SNe
exhibit a broad range for both $\epsilon_{\rm B}$ and $\epsilon_{\rm e}$. From detailed radio and X-ray modeling of SN 2002ap, \citet{bjo04} obtained that the energy density in the electric field is comparable to that of magnetic fields. This was not true for SN 1993J for which \citet{fra98} deduced $\epsilon_{\rm B}/\epsilon_{\rm e} \sim$  280, with $\epsilon_{\rm B} \sim $ 0.14 and $\epsilon_{\rm e} \sim 5 \times 10^{-3}$. In case of SN 2011dh it is found that the radio emission can be interpreted by SSA assuming $\epsilon_{\rm B}$ = $\epsilon_{\rm e}$ =  0.1 \citep{sod12,kra12}. However, to explain the X-ray emission from the same radio emitting electron population one requires $\epsilon_{\rm B}$ = 0.01 and $\epsilon_{\rm e}$ =  0.3 \citep{sod12}. Similarly for SN 2013df, \citet{kam16} estimated a much smaller energy density in magnetic fields compared to that in electric fields ($\epsilon_{\rm B}/\epsilon_{\rm e}$ $\sim 5 \times 10^{-3}$) .

We found that for both SNe 2011fe and 2014J the upper limit on $\mdot /v_w$ has a rather strong dependence on $n$, which varies in the range 12-14, compared to that of $n_{\rm ISM}$ (fig.\ref{fig:const_den_mloss_effect}). However, it is clear from fig.\ref{fig:const_den_mloss_effect} that $\epsilon_{\rm B}$ plays the most important role in determining the upper limits on the CSM density ($n_{\rm ISM}$ or $\mdot/v_w$)  around them. 
From particle in cell (PIC) simulations, \citet{cap14b} showed that for shocks with Mach number ($M_A$) up to 100 the magnetic field amplification is proportional to the square root of $M_A$ (see their eqn.~2 and the bottom panel of fig.5). As $u_e \propto M_A^2$, for shocks with $M_A \sim$ 100 $\frac{u_e}{u_B} \approx M_A$. If this is true for shocks with $M_A \sim $ few $\times$ 1000, which is applicable to SN shocks, then this implies that although the magnetic fields could be amplified very efficiently in post shock region, $u_B$ will be much smaller than $u_e$.


\par

In summary, our study shows that the null detection of radio emission from SNe~2011fe and SN~2014J could be due to the fact that $u_B < u_e$ with $\epsilon_{\rm B} < 0.1$ in the post shock region.  
There are evidences of circumbinary material around a few Type Ia SNe (e.g SNe 2002ic, 2005gj, 2006X; \citet{hum03,ald06,pat07}), however, no radio emission has been detected from them \citep{ber03,stock03, sod05, stock06}. This may indicate that $u_B < u_e$ with $\epsilon_{\rm B} < 0.1$ could be a general feature of SN shocks.  
Future detailed periodic radio observations from nearby SNe together with results from PIC simulations for $M_A \sim$ 1000 could shed light on this.

\section{acknowledgments}
We thank Markus Kromer, for proving us with data from the N100 and violent merger models, and Claes-Ingvar Bj\"ornsson, Emily Freeland, Claes Fransson, Eli Livne, Anthony Piro and Francesco Taddia for discussions. This research was supported by the Munich Institute for Astro- and Particle Physics (MIAPP) of the DFG cluster of excellence "Origin and Structure of the Universe", and has made use of the NASA/IPAC Extragalactic Database (NED) which is operated by the Jet Propulsion Laboratory, California Institute of Technology, under contract with the National Aeronautics and Space Administration. 
The European VLBI Network (EVN) is a joint facility of European, Chinese, South African, and other radio astronomy institutes funded by their national research councils.
The National Radio Astronomy Observatory is a facility of the National Science Foundation operated under cooperative agreement by Associated Universities, Inc.
MAPT, RHI, and AA acknowledge support by the Spanish Ministerio de Economia y Competitividad (MINECO) through the grant AYA2015-63939-C2-1-P, cofunded with FEDER funds.


\end{document}